\documentclass{./jfm}
\usepackage{graphicx}

\title{Bounds for Rayleigh--B\'enard convection between free-slip boundaries with an imposed heat flux}
\shorttitle{Bounds for Rayleigh--B\'enard convection between free-slip, fixed flux boundaries}

\author{ Giovanni Fantuzzi\corresp{\email{gf910@ic.ac.uk}} }
\shortauthor{Giovanni Fantuzzi}

\affiliation{Department of Aeronautics, Imperial College London, 
South Kensington Campus, London, SW7 2AZ, United Kingdom}

\usepackage{tikz}				% tikz for sketches
\usetikzlibrary{plotmarks}		% and library

\usepackage{amsmath}
\usepackage{mathtools}
\usepackage[euler]{textgreek}

\usepackage{xcolor}

\newcommand{\RB}{Rayleigh--B\'enard}
\newcommand{\QUINOPT}{{\sc quinopt}}

\renewcommand{\vec}[1]{\boldsymbol{#1}}
\newcommand{\R}{\mbox{\it R}}
\newcommand{\Ra}{\mbox{\it Ra}}
\newcommand{\Nu}{\mbox{\it Nu}}
\newcommand{\norm}[1]{\left\| #1 \right\|}
\newcommand{\abs}[1]{\left\vert #1 \right\vert}

\newcommand{\dz}{\,{\rm d}z}

\newcommand{\dxi}{\,{\rm d}\xi}
\newcommand{\deta}{\,{\rm d}\eta}
\renewcommand{\phi}{\varphi}

\usepackage{etoolbox}
\makeatletter

\newenvironment{Dequation}
  {%
  \def\tagform@##1{%
    \maketag@@@{\makebox[0pt][r]{(\ignorespaces##1\unskip\@@italiccorr)}}}%
  \ignorespaces
  }
  {%
  \def\tagform@##1{\maketag@@@{(\ignorespaces##1\unskip\@@italiccorr)}}%
  \ignorespacesafterend
  }
 
\makeatother 

\usepackage{hyperref}
\hypersetup{colorlinks=true,linkcolor=blue,urlcolor=blue,citecolor=blue}
\usepackage{natbib}
\makeatletter
\patchcmd{\NAT@citex}
  {\@citea\NAT@hyper@{%
     \NAT@nmfmt{\NAT@nm}%
     \hyper@natlinkbreak{\NAT@aysep\NAT@spacechar}{\@citeb\@extra@b@citeb}%
     \NAT@date}}
  {\@citea\NAT@nmfmt{\NAT@nm}%
   \NAT@aysep\NAT@spacechar\NAT@hyper@{\NAT@date}}{}{}

\patchcmd{\NAT@citex}
  {\@citea\NAT@hyper@{%
     \NAT@nmfmt{\NAT@nm}%
     \hyper@natlinkbreak{\NAT@spacechar\NAT@@open\if*#1*\else#1\NAT@spacechar\fi}%
       {\@citeb\@extra@b@citeb}%
     \NAT@date}}
  {\@citea\NAT@nmfmt{\NAT@nm}%
   \NAT@spacechar\NAT@@open\if*#1*\else#1\NAT@spacechar\fi\NAT@hyper@{\NAT@date}}
  {}{}
  
\makeatother

\begin{document}

\maketitle

\begin{abstract}
We prove the first rigorous bound on the heat transfer for three-dimensional Rayleigh--B\'enard convection  of finite-Prandtl-number fluids between free-slip boundaries with an imposed heat flux. Using the auxiliary functional method with a quadratic functional, which is equivalent to the background method, we prove that the Nusselt number {\Nu} is bounded by $\Nu \leq 0.5999 \R^{1/3}$ uniformly in the Prandtl number, where {\R} is the Rayleigh number based on the imposed heat flux. In terms of the Rayleigh number based on the mean vertical temperature drop, {\Ra},  we obtain $\Nu\leq 0.4646 \Ra^{1/2}$. The scaling with Rayleigh number is the same as that of bounds obtained with no-slip isothermal, free-slip isothermal, and no-slip fixed flux boundaries, and numerical optimisation of the bound suggests that it cannot be improved within our bounding framework. Contrary to the two-dimensional case, therefore, the {\Ra}-dependence of rigorous upper bounds on the heat transfer obtained with the background method for three-dimensional Rayleigh--B\'enard convection is insensitive to both the thermal and the velocity boundary conditions. 
\end{abstract}

% Keywords: select from http://journals.cambridge.org/data/relatedlink/jfm-keywords.pdf
%\begin{keywords}
%B\'enard convection, turbulent convection, variational methods.
%\end{keywords}

%%%%%%%%%%%%%%%%%%%%%%%%%%%%%%%%%%%%%%%%%%%%%%%%%%%%%%%%%%%%%%%%%%%%%%%%%%%%%%%
%%%%%%%%%%%%%%%%%%%%%%%%%%%%%%%%%%%%%%%%%%%%%%%%%%%%%%%%%%%%%%%%%%%%%%%%%%%%%%%

%%%%%%%%%%%%%%%%%%%%%%%%%%%%%%%%%%%%%%%%%%%%%%%%%
\section{Introduction}
\label{s:introduction}

{\RB} (RB) convection, the buoyancy-driven motion of a fluid confined between horizontal plates, is a cornerstone of fluid mechanics. Its applications include atmospheric and oceanic physics, astrophysics, and industrial engineering~\citep[Chapter 3]{Lappa2010}, and due to its rich dynamics it has also become a paradigm to investigate pattern formation and nonlinear phenomena~\citep{Ahlers2009}.
%due to its conceptual simplicity one hand, and its rich dynamics on the other~\citep{Ahlers2009}. 

One of the fundamental questions in the study of convection is to which extent the flow enhances the transport of heat across the layer. Precisely, one would like to relate the Nusselt number {\Nu} (the nondimensional measure of the heat transfer enhancement) to the parameters of the fluid and the strength of the thermal forcing. These are described, respectively, by the Prandtl and Rayleigh numbers $\Pran=\nu/\kappa$ and $\Ra=\alpha g h^3\Delta /(\nu\kappa)$, where $\alpha$ is the fluid's thermal expansion coefficient, $\nu$ is its kinematic viscosity, $\kappa$ is its thermal diffusivity, $h$ is the dimensional height of the layer, $g$ is the gravitational acceleration, and $\Delta$ is the average temperature drop across the layer.
It is generally expected that for large Rayleigh numbers the Nusselt number obeys a simple scaling law of the form $\Nu\sim\Pran^{a}\Ra^{b}$. However, different phenomenological arguments predict different scaling exponents in the ranges $-1/4\leq a \leq 1/2$ and $2/7\leq b \leq 1/2$~\citep[see Tables I and II in][]{Ahlers2009}, and the available experimental evidence in the high-{\Ra} regime is controversial~\citep{Ahlers2009}. 

Discrepancies in the measurements are often attributed to differences in the  boundary conditions (BCs) or in the Prandtl number. From the modelling point of view, eight basic configurations of RB convection can be identified depending on the Prandtl number (finite or infinite), the BCs for the fluid's temperature (fixed temperature or fixed flux), and the BCs for its velocity (no-slip or free-slip). 
Two-dimensional simulations~\citep{Johnston2009,Goluskin2014a,VanDerPoel2014} have shown that changing the thermal BCs for given velocity BCs has no quantitative effect on {\Nu}, while replacing no-slip boundaries with free-slip ones can dramatically reduce the heat transfer through the appearance of zonal flows. However, zonal flows have not been observed in three dimensions~\citep{VanDerPoel2014} and how different BCs affect the {\Nu}-{\Ra}-{\Pran} relationship in general remains an open problem. 

In the absence of extensive numerical result for the high-{\Ra} regime in three dimensions, one way to make progress is through rigorous analysis of the equations that ostensibly describe RB convection. A particularly fruitful approach is to use the background method \citep{Doering1992,Doering1994,Doering1996,Constantin1995a} and derive rigorous bounds of the form $\Nu\leq f(\Ra,\Pran)$ for each of the eight configurations described above.  

The no-slip case has been studied extensively. For fluids with finite Prandlt number the bound $\Nu\lesssim \Ra^{1/2}$ holds uniformly in {\Pran} irrespective of the thermal BCs~\citep{Doering1996,Otero2002,Wittenberg2010a,Wittenberg2010}. When $\Pran=\infty$ (and $\Pran\gtrsim \Ra^{1/3}$ with isothermal boundaries), instead, one has $\Nu\lesssim \ell(\Ra)\Ra^{1/3}$, where $\ell(\Ra)$ is a logarithmic correction whose exact form depends on the thermal BCs~\citep{Otto2011,Whitehead2014,Choffrut2016}. 
%Mixed thermal BCs have also been considered to model poorly conducting boundaries, yielding the same bounds as the fixed flux case~\citep{,Whitehead2014}.
%
%These bounds corroborate numerical evidence that at large {\Ra} the thermal BCs do not affect the heat transfer~\citep{Johnston2009,Stevens2011,Goluskin2014a}. 

In contrast, the only bounds available for free-slip velocity BCs are for RB convection between isothermal plates. All identities and estimates used in the no-slip analysis of~\citet{Doering1996} hold also for free-slip boundaries, so one immediately obtains $\Nu\leq \Ra^{1/2}$ at finite {\Pran}. This result can be tightened to $\Nu\leq \Ra^{5/12}$ in two dimensions and at infinite {\Pran} in three dimensions by explicitly taking advantage of both the stress-free and the isothermal BCs~\citep{Whitehead2011,Whitehead2012}. 
%However, with the exception of recent two-dimensional simulations by~\citet{Goluskin2014a}, the effect of the thermal BCs on free-slip boundaries remains unexplored. 

Free-slip conditions pose a challenge for the background method when a constant heat flux $\kappa\beta$, rather than a fixed boundary temperature, is imposed. The reason is that the analysis usually relies on at least one of the temperature and horizontal velocities being fixed at the top and bottom boundaries, which is not the case with free-slip and fixed flux BCs. 
In this short paper we show that such lack of ``boundary control'' for the dynamical fields can be overcome with a simple symmetry argument and thereby prove the first rigorous upper bound on {\Nu} for RB convection between free-slip boundaries with imposed heat flux. 

The exposition is organised as follows. Section~\ref{s:model} reviews the Boussinesq equations used to model the system. We formulate a bounding principle for {\Nu} in \S\ref{s:bound}, and prove our main result in \S\ref{s:proof}. Finally, \S\ref{s:discussion} offers further discussion and conclusive remarks.

%%%%%%%%%%%%%%%%%%%%%%%%%%%%%%%%%%%%%%%%%%%%%%%%%
\section{The model}
\label{s:model}
We model the system using the Boussinesq equations and make all variables nondimensional using $h$, $h/\kappa$, and $h\beta$, respectively,  as the length, time and temperature scales~\citep{Otero2002}. 
%We model the system using the Boussinesq equations and take $h$, $h/\kappa$, and $h\beta$ as the length, time and temperature scales~\citep{Otero2002}. 
The nondimensional velocity 
%$\vec{u}(x,y,z,t) = u(x,y,z,t)\vec{\hat{i}} + v(x,y,z,t)\vec{\hat{j}} + w(x,y,z,t)\vec{\hat{k}}$,
$\vec{u}(x,y,z,t)$, pressure $p(x,y,z,t)$, and perturbations $\theta(x,y,z,t)$ from the conductive temperature profile $T_c = -z$ then satisfy~\citep{Otero2002}
\begin{subequations}
\begin{gather}
\label{e:momentum}
\partial_t \vec{u}
+ (\vec{u} \bcdot \bnabla) \vec{u} 
+ \bnabla p 
= \Pran\,\nabla^2 \vec{u} + \Pran\,\R\,(\theta - z)  \vec{e}_z,
\\
\label{e:incompress}
\bnabla \bcdot \vec{u} = 0,
\\
\label{e:heat}
\partial_t \theta
+ \vec{u} \bcdot \bnabla \theta = \nabla^2 \theta + w,
\end{gather}
\end{subequations}
where  $\vec{e}_z$ is the unit vector in the $z$ direction and $R=\alpha g \beta h^4 / (\nu \kappa)$ is the Rayleigh number based on the imposed boundary heat flux. Note that {\R} is related to the Rayleigh number based on the (unknown) mean temperature drop, {\Ra}, by $\R = \Ra\,\Nu$~\citep{Otero2002}. The domain is periodic in the horizontal ($x$, $y$) directions and the vertical BCs are
\begin{equation}
\label{e:bc}
\partial_z u = \partial_z  v = w = 0, 
\;
\partial_z  \theta = 0
\text{ at } z=0
\text{ and } z=1.
\end{equation}

Since the average vertical heat flux across the layer is fixed to 1 in nondimensional units, convection reduces the mean temperature difference between the top and bottom plates and hence the mean conductive heat flux 
$\overline{\langle -\partial_z T\rangle}={\overline{ 1-\langle \partial_z \theta\rangle }}$ (here and throughout this work overlines denote averages over infinite time, while angle brackets denote volume averages). The Nusselt number---the ratio of the average vertical heat flux and the mean conductive flux---is then given by~\citep[see also][]{Otero2002}
\begin{equation}
\label{e:Nu}
\Nu =  \left({\overline{ 1-\langle \partial_z \theta\rangle }}\right)^{-1}.
\end{equation}
%
%Here and throughout this work overlines denote averages over infinite time, while angle brackets denote volume averages. 
%In other words, the net heat transfer is inversely proportional to the average flux of heat through the boundaries.

%%%%%%%%%%%%%%%%%%%%%%%%%%%%%%%%%%%%%%%%%%%%%%%%%
\section{Upper bound formulation}
\label{s:bound}

%A variational problem that yields an upper bound on the Nusselt number can be formulated using the background method~\citep{Doering1994,Constantin1995a,Doering1996}. In fact, the background method analysis is identical to that by Otero {\it et al.}~\citep{Otero2002} for no-slip, fixed temperature boundaries because all identities used to formulated the bound hold also in the free-slip case. Instead of considering Otero {\it et al.}'s bounding principle as our starting point, however, we will begin from the equations of motion and use the {\it auxiliary functional method}~\citep{Chernyshenko2017} to compute a strictly positive lower bound $L$ on the time average of $1-\langle \partial_z \theta\rangle$. Our analysis gives exactly the same result as the background method, and makes the optimality of the background method analysis of~\citet{Otero2002} much more transparent because no {\it ad-hoc} linear combinations of identities derived from the equations of motions are required.

When  $\R< 120$ conduction is globally asymptotically stable and $\Nu=1$~\citep{Chapman1980,Goluskin2015}. For $\R>120$ convection sets in~\citep{Hurle1967} and we look for a positive lower bound $L$ on $\overline{1-\langle \partial_z \theta\rangle}$, implying $\Nu \leq 1/L$. To find $L$ we use the background method~\citep{Doering1994,Doering1996,Constantin1995a} but we formulate it in the language of the {\it auxiliary functional method}~\citep{Chernyshenko2014a,Chernyshenko2017} because of its conceptual simplicity: it relies on one simple inequality, rather than a seemingly {\it ad hoc} manipulation of the governing equations.

The analysis starts with the observation that any uniformly bounded and differentiable time-dependent functional $\mathcal{V}(t)=\mathcal{V}\{\theta(\bcdot,t),\vec{u}(\bcdot,t)\}$ satisfies $\overline{{\rm d} \mathcal{V} / {\rm d} t}=0$. Consequently, to prove that $\overline{1-\langle\partial_z\theta\rangle} \geq L$ it suffices to show that at any instant in time
\begin{equation}
\label{e:boundingIneq}
\mathcal{S}\{\theta(\bcdot,t),\vec{u}(\bcdot,t)\} := \frac{ {\rm d} \mathcal{V}}{ {\rm d} t} 
+ 1 - \langle \partial_z \theta\rangle 
- L \geq 0.
\end{equation}
Using the ideas outlined by \citet{Chernyshenko2017}, it can be shown that constructing a background temperature field in the ``classical'' background method analysis is equivalent to finding constants $a$, $b$ and $L$ and a function $\phi(z)$ such that~\eqref{e:boundingIneq} holds for
\begin{equation}
\label{e:V}
\mathcal{V}\{\theta(\bcdot,t),\vec{u}(\bcdot,t)\} := 
-\frac{a}{2\Pran\,\R}\langle \abs{\vec{u}}^2 \rangle
-\frac{b}{2}\langle \theta^2 \rangle 
+ \langle \phi\theta \rangle.
\end{equation}
We assume that $\vec{u}$ and $\theta$ are sufficiently regular in time to ensure differentiability of this functional, while uniform boundedness can be proven using estimates similar to those presented in this paper
%\citep[we do not do so explicitly here, but we refer the interested reader to similar discussions in][]{Doering1992,Hagstrom2014}. 
(we do not give a full proof in this work due to space limitations, but outline the argument in appendix~\ref{a:appA}).

The functional $\mathcal{S}\{\theta(\bcdot,t),\vec{u}(\bcdot,t)\}$ corresponding to~\eqref{e:V} can be expressed in terms of $\vec{u}$ and $\theta$ using~\eqref{e:momentum}--\eqref{e:heat}. Integrating the volume average $\langle \vec{u}\bcdot$\eqref{e:momentum}$\rangle$ by parts using incompressibility and the BCs yields
\begin{equation}
\label{e:part2}
\frac{\rm d}{{\rm d} t}
\frac{\langle \abs{\vec{u}}^2 \rangle}{2\Pran\,\R} = 
-\frac{\langle \abs{\bnabla\vec{u}}^2 \rangle}{\R} 
+ \langle w \theta \rangle.
\end{equation}
Averaging $\theta\times$\eqref{e:heat} and $\phi\times$\eqref{e:heat} in a similar way gives
\begin{align}
\label{e:part3}
\frac{\rm d}{{\rm d} t}\frac{\langle \theta^2 \rangle}{2} &=  - \langle \abs{\bnabla\theta}^2 \rangle + \langle w\theta \rangle,
\\
\label{e:part4}
\langle \phi\,\partial_t \theta \rangle &= 
\langle \phi' w\theta \rangle
-\langle \phi' \partial_z \theta \rangle.
\end{align}
Combining expressions~\eqref{e:part2}--\eqref{e:part4} and rearranging we find
\begin{equation}
\label{e:S}
\mathcal{S}\{\theta(\bcdot,t),\vec{u}(\bcdot,t)\} = 1 - L 
- \left\langle(\phi'+1)\partial_z\theta \right\rangle
+ \left\langle 
\frac{a}{\R} \abs{\bnabla\vec{u}}^2 + b\abs{\bnabla\theta}^2 + (\phi'-a-b)w\theta
\right\rangle.
\end{equation}

To prove that~\eqref{e:boundingIneq} holds at all times, we make one key further simplification: we drop the equation of motions and choose $a$, $b$, $L$ and $\phi(z)$ such that $\mathcal{S}\{\theta,\vec{u}\}\geq 0$ for any time-independent fields $\theta = \theta(x,y,z)$ and $\vec{u}= \vec{u}(x,y,z)$ that satisfy~\eqref{e:incompress} and the BCs. Hereafter, we also assume that $a,b>0$ to ensure that $\mathcal{S}\{\theta,\vec{u}\}$ is bounded below. 

Incompressibility can be incorporated explicitly in~\eqref{e:S} upon substitution of the horizontal Fourier expansions
%%%%%%%%%%%%%%%
%% Must do this manually otherwise get wrong numbers???
\stepcounter{equation}
%%%%%%%%%%%%%%%
% Dequation only works with align?
\begin{Dequation}
\begin{align}
\label{e:FourierVars}
\tag{\theequation{\it a,b}}
\theta = \sum_{\vec{k}} \theta_{\vec{k}}(z) {\rm e}^{\vec{k}\bcdot\vec{x}},
\qquad
\vec{u} = \sum_{\vec{k}} \vec{u}_{\vec{k}}(z) {\rm e}^{\vec{k}\bcdot\vec{x}},
\end{align}
\end{Dequation}
where $\vec{x}=(x,y)$ is the horizontal position vector and $\vec{k}=(k_x,k_y)$ is the wavevector. The $z$-dependent Fourier amplitudes $\vec{u}_{\vec{k}}$, $\theta_{\vec{k}}$ satisfy the same vertical BCs as the full fields in~\eqref{e:bc}. %After changing variables from $u_{\vec{k}}$ and $v_{\vec{k}}$ to $u_{\vec{k}}\cos\omega_{\vec{k}}+ v_{\vec{k}}\sin\omega_{\vec{k}}$ and $v_{\vec{k}}\cos\omega_{\vec{k}}- u_{\vec{k}}\sin\omega_{\vec{k}}$ with $\omega_{\vec{k}} = \tan^{-1}(k_y/k_x)$ (and $\omega_{\vec{k}}=\pi/2$ when $k_x=0$)~\citep{Doering1996}, 
Using the Fourier-transformed incompressibility constraint one can show that~\citep{Doering1996,Otero2002}
\begin{equation}
\label{e:Sfourier}
\mathcal{S}\{\theta,\vec{u}\} \geq 
\mathcal{S}_0\{\theta_0\} 
+ b \sum_{\vec{k}\neq (0,0)} 
\mathcal{S}_{\vec{k}}\{\theta_{\vec{k}},w_{\vec{k}}\},
\end{equation}
with
\begin{subequations}
\begin{equation}
\label{e:S0}
\mathcal{S}_0\{\theta_0\} \!:=
b\norm{\theta_0'}^2 
-\int_0^1\!(\phi'+1)\theta_0' \dz + 1-L,
\end{equation}
\begin{multline}
\label{e:Sk}
\mathcal{S}_{\vec{k}} \{\theta_{\vec{k}},w_{\vec{k}}\} \!:= 
\norm{\theta_{\vec{k}}'}^2 + k^2 \norm{\theta_{\vec{k}}}^2 
+\frac{a}{b\R}\left( 
\frac{\norm{w_{\vec{k}}''}^2}{k^2} 
+ 2\norm{w_{\vec{k}}'}^2
+ k^2 \norm{w_{\vec{k}}}^2\right)
\\
+\int_0^1\frac{\phi'-a-b}{b}\,{\rm Re}( \theta_{\vec{k}} \tilde{w}_{\vec{k}}) \dz.
\end{multline}
\end{subequations}
In these equations and in the following we write $k^2 = k_x^2 + k_y^2$, $\norm{\bcdot}$ denotes the standard Lebesgue $\mathcal{L}^2$ norm on the interval $(0,1)$, and $\tilde{w}_{\vec{k}}$ is the complex conjugate of $w_{\vec{k}}$.

The right-hand side of~\eqref{e:Sfourier} is clearly non-negative if $\mathcal{S}_0\geq 0$ and $\mathcal{S}_{\vec{k}}\geq 0$ for all wavevectors $\vec{k} \neq (0,0)$. (A standard argument based on the consideration of fields $\theta$ and $\vec{u}$ with a single Fourier mode shows that these conditions are also necessary, so enforcing the positivity of each $\mathcal{S}_{\vec{k}}$ individually does not introduce conservativeness. However, necessity is not required to proceed with our argument so we omit the details for brevity.)
In particular, given  $a$, $b$, and $\phi$ the largest value of $L$ for which $\mathcal{S}_0\geq 0$ is found upon completing the square (in the $\mathcal{L}^2$ norm sense) in~\eqref{e:S0}, so we set
\begin{equation}
\label{e:Lopt0}
L = 1 - \frac{\norm{\phi' + 1}^2}{4b}.
\end{equation}
We will try to maximise this expression over $a$, $b$ and $\phi$ subject to the non-negativity of the functional $\mathcal{S}_{\vec{k}}$ in~\eqref{e:Sk} for all wavevectors $\vec{k}\neq (0,0)$. Note that  $\mathcal{S}_{\vec{k}}$ and the right-hand side of~\eqref{e:Lopt0}  reduce, respectively, to the quadratic form and the bound obtained by~\citet{Otero2002} using the ``classical'' background method analysis if we let $a=b-1$ and identify $[\phi'(z)-2b+1]/(2b)$ with the derivative of the background temperature field. We also remark that our analysis appears more general because the choice $a=1-b$ is unjustified at this stage, but its optimality (at least within the context of our proof) will be demonstrated below.

%%%%%%%%%%%%%%%%%%%%%%%%%%%%%%%%%%%%%%%%%%%%%%%%%
\section{An explicit bound}
\label{s:proof}

\begin{figure}
\centering
\includegraphics[scale=0.9,trim=0.1cm 0.25cm 0.1cm 0cm]{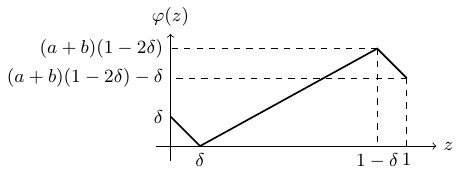}
\caption{\label{f:phi}Sketch of the piecewise-linear function $\phi(z)$.}
\end{figure}

Let $\delta\leq 1/2$ and consider the piecewise-linear profile $\phi(z)$ shown in figure~\ref{f:phi}, whose derivative is
\begin{equation}
\label{e:phi}
\phi'(z) = \begin{cases}
\phantom{a}-1, & z\in[0,\delta]\cup[1-\delta,1],\\
a+b, &z\in(\delta,1-\delta).
\end{cases}
\end{equation}
To show that $a$, $b$, and $\delta$ can be chosen to make the quadratic form $\mathcal{S}_{\vec{k}} \{\theta_{\vec{k}},w_{\vec{k}}\}$ in~\eqref{e:Sk} positive semidefinite we rewrite $\theta_{\vec{k}}$ and $w_{\vec{k}}$ as the sum of functions that are symmetric and antisymmetric with respect to $z=1/2$. In other words, we decompose
%
%%%%%%%%%%%%%%%
%% Must do this manually otherwise get wrong numbers???
\stepcounter{equation}
%%%%%%%%%%%%%%%
% Dequation only works with align?
\begin{Dequation}
\begin{align}
\tag{\theequation{\it a,b}}
\theta_{\vec{k}}(z) = {\theta}_+(z) + {\theta}_-(z),
\quad
w_{\vec{k}}(z) = {w}_+(z) + {w}_-(z),
\end{align}
\end{Dequation}
with
%
%%%%%%%%%%%%%%%
%% Must do this manually otherwise get wrong numbers???
\stepcounter{equation}
%%%%%%%%%%%%%%%
% Dequation only works with align?
\begin{Dequation}
\begin{align}
\tag{\theequation{\it a,b}}
\theta_\pm(z) = \frac{\theta_{\vec{k}}(z) \pm \theta_{\vec{k}}(1-z)}{2},
\qquad
w_\pm(z) = \frac{w_{\vec{k}}(z) \pm w_{\vec{k}}(1-z)}{2}.
\end{align}
\end{Dequation}
(The subscripts $+$ and $-$ denote, respectively, the symmetric and antisymmetric parts.) Since $\phi'(z)$ is symmetric with respect to $z=1/2$ by construction we obtain
\begin{equation}
\label{e:evenoddSk}
\mathcal{S}_{\vec{k}} \{\theta_{\vec{k}},w_{\vec{k}}\} = 
\mathcal{S}_{\vec{k}}\{ \theta_+,w_+\} 
+ \mathcal{S}_{\vec{k}} \{ \theta_-,w_-\},
\end{equation}
{\it i.e.}, we can split the quadratic form $\mathcal{S}_{\vec{k}} \{\theta_{\vec{k}},w_{\vec{k}}\}$ into its symmetric and antisymmetric components also.
Symmetric and antisymmetric Fourier amplitudes $\theta_{\vec{k}}$ and $w_{\vec{k}}$---for which one term on the right-hand side of~\eqref{e:evenoddSk} vanishes---are also admissible, so $\mathcal{S}_{\vec{k}}$ is non-negative if and only if it is so for arguments that are either symmetric or antisymmetric with respect to $z=1/2$ (and, of course, satisfy the correct BCs). As before, the ``only if'' statement is not needed to proceed but guarantees that no conservativeness is introduced.
% Needed to avoind ugly space?
%\vskip -2.75ex

%%%%%%%%%%%%%%%%%%%%%%%%%%%
%%%%%%%%%%%%%%%%%%%%%%%%%%%%

The decomposition into symmetric and antisymmetric components is the essential ingredient of our proof.  In fact, contrary to the case of no-slip boundaries considered by~\citet{Otero2002}, the free-slip and fixed flux BCs cannot be used to control the indefinite term in $\mathcal{S}_{\vec{k}}\{ \theta_{\vec{k}},w_{\vec{k}}\}$ via the usual elementary functional-analytic estimates. However, $\theta_\pm$ and $w_\pm$ (or the appropriate derivatives) are known not only at the boundaries, but also on the symmetry plane. 
%In particular, the conditions
%%
%\begin{equation}
%\label{e:symmbc}
%w_\pm(0)= w_+'(1/2) = \theta_-(1/2) = 0
%\end{equation}
%%
%give control of the indefinite term in~\eqref{e:Sk} when $\delta$ is sufficiently small.
In particular, for small $\delta$ the indefinite term in~\eqref{e:Sk} can be controlled without recourse to the free-slip, fixed-flux BCs using
\begin{equation}
\label{e:symmbc}
w_\pm(0)= w_+'(1/2) = \theta_-(1/2) = 0.
\end{equation}
%
% [the exact value will be given in~\eqref{e:delta}]. 

To prove this, recall that for any symmetric or antisymmetric quantity $q(z)$
\begin{equation}
\label{e:normequiv}
\int_0^{1/2} \abs{q(z)}^2 \dz 
= \frac{\norm{q}^2}{2}.
\end{equation}
Symmetry,~\eqref{e:phi}, and the identity $\abs{\theta_\pm \tilde{w}_\pm} = \abs{\theta_\pm {w}_\pm}$ ($\tilde{w}_\pm$ is the complex conjugate of ${w}_\pm$) yield
\begin{equation}
\label{e:est0}
\abs{\int_0^1\!
\frac{\phi'-a-b}{b}
{\rm Re}(\theta_\pm \tilde{w}_\pm)
\!\dz} 
\leq 2M\!\int_0^\delta\!\abs{ \theta_\pm {w}_\pm} \dz,
\end{equation}
with $M := (1+a+b)/b$. 
Since $w_\pm(0)=0$ the product $\theta_\pm w_\pm$ vanishes at $z=0$ and for any $z\leq \delta \leq 1/2$ the fundamental theorem of calculus implies
\begin{equation}
\label{e:est1}
\abs{\theta_\pm(z) {w}_\pm(z)} \leq 
\int_0^z \abs{\theta_\pm(\xi)}\abs{w_\pm'(\xi)}\dxi +
\int_0^z \abs{\theta_\pm'(\xi)}\abs{w_\pm(\xi)}\dxi.
\end{equation}
Using the fact that $w_\pm(0)=0$ once again, the  fundamental theorem of calculus for $\xi\leq 1/2$, the Cauchy--Schwarz inequality, and~\eqref{e:normequiv} we also obtain
\begin{equation}
\label{e:est2}
\abs{w_\pm(\xi)} = \abs{\int_0^{\xi} w_\pm'(\eta) \deta} 
\leq \sqrt{\frac{\xi}{2}}\norm{w_\pm'}.
\end{equation}
Furthermore, the conditions in~\eqref{e:symmbc} imply that the product $\theta_\pm w_\pm'$ vanishes at the symmetry plane, so similar estimates as above yield
\begin{align}
\label{e:est3}
\abs{\theta_\pm(\xi)w_\pm'(\xi)} &= \abs{\int_\xi^{1/2} \left[\theta_\pm(\eta)w_\pm''(\eta) + \theta_\pm'(\eta)w_\pm'(\eta) \right]\!\deta} 
\notag
\\&\leq 
\frac{1}{2}\norm{\theta_\pm}\norm{w_\pm''} + \frac{1}{2}\norm{\theta_\pm'}\norm{w_\pm'}.
\end{align}
Upon inserting~\eqref{e:est2} and~\eqref{e:est3} into~\eqref{e:est1}, applying the Cauchy--Schwarz inequality, and using~\eqref{e:normequiv} we arrive at
\begin{equation}
\abs{\theta_\pm(z) {w}_\pm(z)} \leq 
\frac{z}{2} \left( \norm{\theta_\pm} \norm{w_\pm''} + \frac{1+\sqrt{2}}{\sqrt{2}} \norm{\theta_\pm'} \norm{w_\pm'}\right).
\end{equation}
Substituting this estimate into~\eqref{e:est0} and integrating gives an estimate for the indefinite term in~\eqref{e:Sk}, and after dropping the term $ak^2\norm{w_\pm}^2/(b\R)$ we conclude that
\begin{multline}
\label{e:Slb}
\mathcal{S}_{\vec{k}}\{\theta_\pm,w_\pm\} \geq\;
\frac{2a}{b\R} \norm{w_\pm'}^2
- \frac{(1+\sqrt{2})M\delta^2}{2\sqrt{2}}\norm{w_\pm'}\norm{\theta_\pm'}
+ \norm{\theta_\pm'}^2
\\
+\frac{a}{b\R k^2} \norm{w_\pm''}^2 
- \frac{M\delta^2}{2}\norm{w_\pm''}\norm{\theta_\pm}
+ k^2 \norm{\theta_\pm}^2
\end{multline}
Recalling the definition of $M$ and that a quadratic form $\alpha u^2 + \beta uv + \gamma v^2$ is positive semidefinite if $\beta^2 \leq 4\alpha\gamma$, the right-hand side of~\eqref{e:Slb} is non-negative if we set
%%%%%%%%%%%%%%%
%% Must do this manually otherwise get wrong numbers???
\stepcounter{equation}
%%%%%%%%%%%%%%%
% Dequation only works with align?
\begin{Dequation}
\begin{align}
\label{e:delta}
\delta = A \left[\frac{ab}{(1+a+b)^2\R}\right]^{1/4}, 
\qquad A:=\left(  \frac{8}{1+\sqrt{2}} \right)^{1/2}.
\tag{\theequation{\it a,b}}
\end{align}
\end{Dequation}

Having chosen $\delta$ to ensure the non-negativity of $\mathcal{S}_{\vec{k}}$, all is left to do is optimise the eventual bound $\Nu\leq L^{-1}$ over $a$ and $b$ as a function of {\R}. Substituting~\eqref{e:phi} into~\eqref{e:Lopt0} for our choice of  $\delta$ yields
\begin{equation}
\label{e:Lopt2}
L = 1 -\frac{(1+a+b)^2}{4b}
+\frac{A}{2}\frac{ (1+a+b)^{3/2}a^{1/4}}{ b^{3/4} R^{1/4}}.
\end{equation}
In order to maximise this expression with respect to $a,b>0$ we set the partial derivatives $\partial L/\partial a$ and $\partial L/\partial b$ to zero. After some rearrangement it can be verified that
\begin{subequations}
\begin{align}
\frac{\partial L}{\partial a} &= 0 \quad \Leftrightarrow \quad
A b^{1/4}(7a+b+1)-4 R^{1/4}a^{3/4}(1+a+b)^{1/2} = 0,\\
\label{e:stationary2}
\frac{\partial L}{\partial b} &= 0 \quad \Leftrightarrow \quad
(1+a-b)\left[ 2 R^{1/4} (1+a+b)^{1/2} - 3A a^{1/4} b^{1/4}\right] =0.
\end{align}
\end{subequations}
A few lines of simple algebra show that setting to zero the second factor in~\eqref{e:stationary2} leads to a solution with negative $a$ or $b$, so we must choose $b=1+a$ where $a>0$ satisfies
\begin{equation}
\label{e:xieqn}
A^4(1+4a)^4-64\R(1+a)a^3=0.
\end{equation}
(No positive roots exist if $R \leq 4 A^4 \approx 43.92$, but we are only interested in $\R\geq 120$ because conduction is globally asymptotically stable otherwise. 
It can also be checked that this stationary point is a maximum; the algebra is straightforward but lengthy and uninteresting, so we do not report it for brevity). 
In particular, when {\R} tends to infinity~\eqref{e:xieqn} admits an asymptotic solution of the form $a = a_1\R^{-1/3} + O(\R^{-2/3})$.
Substituting this expansion into~\eqref{e:xieqn} and solving for the leading order terms gives $a_1 = A^{4/3}/4$. We then set $b=1+a$ and $a=a_1 \R^{-1/3}$ in~\eqref{e:Lopt2}, simplify, and estimate
\begin{equation}
L = \frac{A^{4/3}}{4\R^{1/3}}\left[ 
\sqrt{2}\left(4+\frac{A^{4/3}}{\R^{1/3}}\right)^{3/4}
-1 \right]
\geq \frac{3A^{4/3}}{4\R^{1/3}}.
\end{equation}
Note that this bound is sharp as $\R\to\infty$. Consequently,
\begin{equation}
\label{e:AnalyticalBound}
\Nu \leq \frac{1}{L} \leq \frac{4\R^{1/3}}{3A^{4/3}}  \approx 0.5999 \, \R^{1/3}.
\end{equation}
Recalling that $\R=\Nu\,\Ra$~\citep{Otero2002} we can also express this bound in terms of the Rayleigh number {\Ra} based on the average temperature drop across the layer:
\begin{equation}
\label{e:BoundRa}
\Nu \leq %(8\Ra^{1/2})/(3\sqrt{3}A^{2})  
\frac{ 8\Ra^{1/2} }{ 3\sqrt{3}A^{2} }
\approx 0.4646\,\Ra^{1/2}.
\end{equation}

%%%%%%%%%%%%%%%%%%%%%%%%%%%%%%%%%%%%%%%%%%%%%%%%%
\section{Discussion}
\label{s:discussion}

The bound proven in this work is the first rigorous result for three-dimensional RB convection between free-slip, fixed flux boundaries (but note that our proof holds also in the two-dimensional case). Key to the result is a symmetry argument that overcomes the loss of boundary control for the trial fields when the no-slip velocity conditions are replaced with free-slip ones. Our approach is fully equivalent to the ``classical'' application of the background method to the temperature field, and the scaling of our bound with {\Ra} is the same as obtained for no-slip BCs~\citep[irrespectively of the thermal BCs, see][]{Doering1996,Otero2002} and for free-slip isothermal BCs~\citep{Doering1996}. Modulo differences in the prefactor, therefore, rigorous upper bounds on the the heat transfer obtained with the background method for three-dimensional RB convection at finite {\Pran} are insensitive to both the velocity and the thermal BCs.

Whether convective flows observed in reality exhibit the same lack of sensitivity to the BCs, however, remains uncertain. Two-dimensional simulations indicate that the thermal BCs make no quantitative difference for given velocity BCs~\citep{Johnston2009,Goluskin2014a}, while replacing no-slip with free-slip leads to zonal flows with reduced vertical heat transfer~\citep{Goluskin2014a,VanDerPoel2014}. Partial support for such observations comes from the improved bound $\Nu \lesssim \Ra^{5/12}$ obtained with free-slip isothermal boundaries in two dimensions~\citep{Whitehead2011}. It does not seem unreasonable to expect that a symmetry argument similar to that of this paper will extend the result to the fixed flux case, but we leave a formal confirmation to future work.
On the other hand, zonal flows have not been observed in three dimensions~\citep{VanDerPoel2014}. More extensive three-dimensional numerical simulations should be carried out to reveal if and how free-slip conditions affect the {\Nu}-{\Ra} relationship, as well as whether the thermal BCs can have any influence.

\begin{figure}
\centering
\includegraphics[scale=1,trim=0cm 0cm 0cm 0cm]{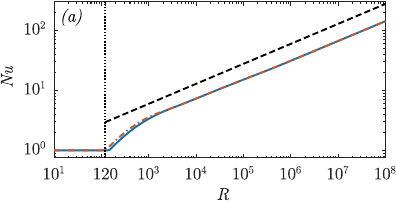}
\hfill%\hspace{1em}
\includegraphics[scale=1,trim=0cm 0cm 0cm 0cm]{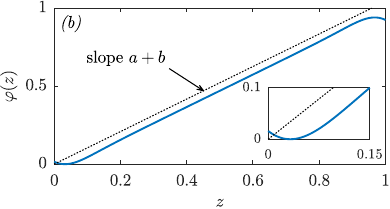}
\caption{\label{f:bound}
(Colour online) 
{\it (a)} The analytic bound~\eqref{e:AnalyticalBound} (dashed line) vs. the numerically optimal bounds for a $2\pi$-periodic layer (solid line) and a $10\pi$-periodic layer (dot-dashed line). The vertical dotted line at $R=120$ marks the global stability boundary for pure conduction.
{\it (b)} The optimal $\phi(z)$ at $\R=10^5$. A dotted line with slope $a+b$ for the optimal values of these parameters at $\R=10^5$ is shown for comparison with the analytical profile sketched in figure~\ref{f:phi}.
}
\end{figure}

%Should {\Nu} be expected to grow more slowly than $\Ra^{1/2}$, however, to confirm so rigorously by means of upper bounds will require going beyond the classical background method analysis. 
Should numerical simulations in three dimensions suggest that {\Nu} grows more slowly than $\Ra^{1/2}$, the challenge will be to improve the scaling exponents in~\eqref{e:AnalyticalBound}--\eqref{e:BoundRa}. The argument by~\citet{Whitehead2012} may be adapted to study the infinite-{\Pran} limit, but cannot be used at finite {\Pran}. Moreover, at finite {\Pran} it does not seem sufficient to consider a more sophisticated choice of $a$, $b$, and $\phi(z)$  in the functional~\eqref{e:V}. To provide evidence of this fact, we used {\QUINOPT}~\citep{Fantuzzi2017tac,quinopt} to maximise the constant $L$ (and, consequently, minimise the eventual bound $\Nu\leq L^{-1}$) over all constants $a$, $b$ and functions $\phi(z)$ that make the functionals in~\eqref{e:S0} and~\eqref{e:Sk} positive semidefinite. We considered domains with period $2\pi$ and $10\pi$ in both horizontal directions, respectively, and the corresponding optimal bounds on {\Nu} are compared to the analytic bound~\eqref{e:AnalyticalBound} in figure~\ref{f:bound}{\it(a)}. For both values of the horizontal period a least-square power-law fit to the numerical results for $\R\geq 10^6$ returns $L^{-1}\approx 0.325\,\R^{0.33}$. Moreover, as illustrated in figure~\ref{f:bound}{\it(b)} for $\R=10^5$, the optimal $\phi(z)$ closely resembles the analytical profile sketched in figure~\ref{f:phi}: it is approximately linear with slope $a+b$ in the bulk and it decreases near the top and bottom boundaries.  This strongly suggests that carefully tuning $a$, $b$, and $\phi(z)$ can only improve the prefactor in~\eqref{e:AnalyticalBound}. 

Lowering the scaling exponent for three-dimensional RB convection at finite Prandtl number, if at all possible, will therefore demand a different approach. Recently, \Citet{Tobasco2017arxiv} have proven that the auxiliary functional method gives arbitrarily sharp bounds on maximal time averages for systems governed by ordinary differential equations. This gives hope that progress may be achieved in the context of RB convection if a more general functional than~\eqref{e:V} is considered. The resulting bounding problem will inevitably be harder to tackle with purely analytical techniques,  but the viability of this approach may be assessed with computer-assisted investigation based on sum-of-squares programming
\citep[see, e.g.,][]{Goulart2012,Chernyshenko2014a,Fantuzzi2016siads,Goluskin2016}. Another option is to try and lower the bound proven here through the study of optimal ``wall-to-wall'' transport problems~\citep{Hassanzadeh2014,Tobasco2017}. Exactly how much these alternative bounding techniques can improve on the background method and advance our ability to derive a rigorous quantitative description of hydrodynamic systems is the subject of ongoing research.

% may improve the scaling of the bound proven in this paper for the general three-dimensional arbitrary-{\Pran}. However, incompressible flows that transport heat to within logarithms of the rigorous upper bounds have recently been constructed for no-slip, isothermal boundaries~\citep{Tobasco2017}, and it does not seem unreasonable to expect that similar ones exist for the free-slip, fixed flux case. Such flows would become admissible trial fields for the auxiliary functional bounding framework whenever the equation of motions are disregarded, irrespective of the choice of auxiliary functional. Consequently, it seems likely that any progress (if at all possible) will require exploiting the dynamic coupling between temperature and velocity.

%%%%%%%%%%%%%%%%%%%%%%%%%%%%%%%%%%%%%%%%%%%%%%%%%%%%%%%%%%%%%%%%%%%%%%%%%%
% Acknowledgements
\vspace{2ex}
We are indebted to D.~Goluskin and J.~P.~Whitehead, who introduced us to the problem studied in this paper. We thank them, C.~R.~Doering, A.~Wynn, and S.~I.~Chernyshenko for their encouragement and helpful comments. Funding by an EPSRC scholarship (award ref. 1864077) and the support and hospitality of the Geophysical Fluid Dynamics program at Woods Hole Oceanographic Institution are gratefully acknowledged. 

%%%%%%%%%%%%%%%%%%%%%%%%%%%%%%%%%%%%%%%%%%%%%%%%%%%%%%%%%%%%%%%%%%%%%%%%%%
%% Appendix
\appendix
\section{Boundedness of $\mathcal{V}$}\label{a:appA}
The Cauchy-Schwarz inequality and the estimate $\langle \abs{\theta}^2\rangle = \langle \abs{T + z}^2\rangle\leq 2\langle \abs{T}^2\rangle + 2/3$ imply that the functional in~\eqref{e:V} is bounded if $\langle \abs{\vec{u}}^2\rangle,\langle \abs{T}^2\rangle<\infty$. Following ideas by \citet{Doering1992} and \citet{Hagstrom2014}, this holds if velocity and temperature perturbations $\hat{\vec{u}}:= \vec{u}-\vec{\psi}$ and $\vartheta:=T-\tau$ from steady background fields $\vec{\psi}$ and $\tau$ satisfy $\langle \abs{\hat{\vec{u}}}^2\rangle,\langle \abs{\vartheta}^2\rangle<\infty$. Below we briefly outline how to find suitable $\vec{\psi}$ and $\tau$. 

Let $T(\cdot, 0)$ and $\vec{u}(\cdot, 0)$ be given initial conditions. The volume-averaged temperature  and horizontal velocities ($\langle T\rangle$, $\langle u\rangle$ and $\langle v\rangle$) are conserved, {\it e.g.} $\langle T(\cdot,t)\rangle = \langle T(\cdot, 0)\rangle$. This follows after taking the volume average of the Boussinesq equations using the divergence theorem, incompressibility, and the BCs. Then, let $\vec{\psi}:=\langle u(\cdot, 0)\rangle\vec{e}_x+\langle v(\cdot, 0)\rangle\vec{e}_y$ and set $\tau=\tau(z)$ with
\begin{equation}
\label{e:tau}
\tau'(z) = \begin{cases}
-1, &z\in[0,\delta]\cup[1-\delta,1],\\
\phantom{-}1, &z\in(\delta,1-\delta),
\end{cases}
\end{equation}
for some $\delta>0$ to be determined and the constant of integration chosen such that $\int_0^1 \tau(z) \dz = \langle T(\cdot, 0)\rangle$. It follows that $\hat{\vec{u}}$ satisfies the same BCs as the full velocity field, $\vartheta$ satisfies $\partial_z\vartheta\vert_{z=0}=0=\partial_z\vartheta\vert_{z=1}$, and $\langle \vartheta \rangle=\langle \hat{u} \rangle=\langle \hat{v} \rangle=0$ at all times.

Since $\vec{\psi}$ and $\tau$ are independent of time and $\bnabla\bcdot\vec{\psi}=0$, for any constant $C>0$ we can use  incompressibility, the BCs, and the Boussinesq equations to write
\begin{equation}
\label{e:A1}
\frac{{\rm d }}{{\rm d} t}\left\langle \frac{\abs{\vartheta}^2}{2} + \frac{\abs{\hat{\vec{u}}}^2}{2\Pran\,\R}\right\rangle  = 
- \left\langle 
\abs{\bnabla \vartheta}^2 + \frac{\abs{\bnabla\hat{\vec{u}}}^2}{\R} 
+ (\tau'-1) \hat{w}\vartheta 
+(\tau'+1)\partial_z \vartheta 
+C\right\rangle + C.
\end{equation}
The task is then to find $\delta$ in~\eqref{e:tau}, $C>0$, and a constant $\gamma>0$ such that
\begin{equation}
\label{e:A2}
\left\langle 
\abs{\bnabla \vartheta}^2 + \frac{\abs{\bnabla\hat{\vec{u}}}^2}{\R} 
+ (\tau'-1) \hat{w}\vartheta 
+(\tau'+1)\partial_z \vartheta 
+C\right\rangle - \gamma\left\langle \frac{\abs{\vartheta}^2}{2} + \frac{\abs{\hat{\vec{u}}}^2}{2\Pran\,\R}\right\rangle\geq 0
\end{equation}
for all time-independent trial fields $\hat{\vec{u}}$ and $\vartheta$ with $\langle \vartheta \rangle=\langle \hat{u} \rangle=\langle \hat{v} \rangle=0$ and $\bnabla\bcdot\hat{\vec{u}}=0$ that satisfy the BCs. In fact, combining~\eqref{e:A1} and~\eqref{e:A2} shows that $\langle \abs{\vartheta}^2/2 + \abs{\hat{\vec{u}}}^2/(2\Pran\,\R)\rangle$ decays when it is large, remaining bounded. Hence, $\langle \abs{\hat{\vec{u}}}^2\rangle$ and $\langle \abs{\vartheta}^2\rangle$ are also bounded.

Inequality~\eqref{e:A2} can be proven wavenumber by wavenumber upon considering horizontal Fourier expansions for $\vartheta$ and $\hat{\vec{u}}$ provided that {\it(i)} $\gamma< \min\{4,\,4\Pran,\,2k_m^2,\,2\Pran k_m^2\}$ with $k_m^2 := \min_{\vec{k}\neq(0,0)}k^2$ (here $k^2$ is the magnitude of the horizontal wavevector, cf. \S\ref{s:bound}; the minimum is strictly positive because we work in a finite periodic domain), {\it(ii)} $C$ is sufficiently large, and {\it(iii)} $\delta$ is sufficiently small. Nonzero wavevectors can be analysed using estimates similar to those of \S\ref{s:proof}, while the case $\vec{k}=(0,0)$ is handled using Poincar\'e-type inequalities deduced using the zero-average conditions $\langle \vartheta \rangle=\langle \hat{u} \rangle=\langle \hat{v} \rangle=0$.
\bibliographystyle{./jfm}
\bibliography{refs}

\end{document}